\documentclass[prd,twocolumn,nofootinbib,letterpaper,floatfix]{revtex4}
\usepackage{color}

\usepackage{graphicx}

\usepackage{amsmath}
\usepackage{amssymb}
\usepackage{subfigure}
\usepackage{url}

\usepackage{multirow}
\usepackage{threeparttable}
\usepackage{paralist}

\newcommand{\GeV}{\text{~GeV}}
\newcommand{\TeV}{\text{~TeV}}

\DeclareRobustCommand{\Sec}[1]{Sec.~\ref{#1}}

\DeclareRobustCommand{\Tab}[1]{Table~\ref{#1}}

\DeclareRobustCommand{\Fig}[1]{Fig.~\ref{#1}}

\DeclareRobustCommand{\Eq}[1]{Eq.~(\ref{#1})}

\newcommand{\be}{\begin{equation}}
\newcommand{\ee}{\end{equation}}

\newcommand{\mb}[1]{\boldsymbol{#1}}
\newcommand{\bm}[1]{\boldsymbol{#1}}

\newcommand{\kahler}{K\"{a}hler }

\newcommand{\rep}[1]{\mathbf{#1}}

\renewcommand{\a}{\alpha}

\newcommand{\D}{\Delta}
\renewcommand{\l}{\lambda}
\renewcommand{\th}{\theta}
\newcommand{\IZ}{\mathbb{Z}}

\newcommand{\cL}{\mathcal{L}}

\begin{document}

\title{Small $m_\nu$ due to $R$-symmetry breaking for small $\Lambda$}

\author{Rhys Davies}
\email{daviesr@maths.ox.ac.uk}
\affiliation{Mathematical Institute, University of Oxford, 24-29 St Giles, Oxford OX1 3LB, UK}

\author{Matthew McCullough}
\email{mccull@mit.edu}
\affiliation{Center for Theoretical Physics, Massachusetts Institute of Technology, Cambridge, MA 02139, USA}

\date{\today}

\begin{abstract}
We describe a class of supersymmetric models in which neutrinos are kept
light by an $R$-symmetry.  In supergravity, $R$-symmetry must be broken
to allow for a small cosmological constant after supersymmetry breaking.  In
the class of models described here, this $R$-symmetry breaking results in
the generation of Dirac neutrino masses, connecting the tuning of the cosmological 
constant to the puzzle of neutrino masses.  Surprisingly, under the assumption of low-scale supersymmetry breaking and superpartner masses close to a TeV, these masses are
independent of the fundamental supersymmetry-breaking scale, and accommodate the
correct magnitude.   This offers a novel explanation for the vastly different
scales of neutrino and charged fermion masses.  These models require that
$R$-symmetric supersymmetry exists at the TeV scale, and predict that
neutrino masses are purely Dirac, implying the absence of neutrino-less double beta-decay.  Interesting collider signals can arise 
due to charged scalars which decay leptonically, with branching ratios determined 
by the neutrino mixing matrix.

\end{abstract}

\preprint{MIT-CTP 4325}

\maketitle

\section{Introduction}\label{sec:introduction}
Although it has now been established that at least two generations of
neutrinos have mass, the origin and nature of their mass is still
undetermined.  The magnitude of neutrino masses lies well below the
weak scale, but it is compelling and economical to consider whether
they may be related to the physics that determines, or at least
stabilizes, the weak scale.  A number of suggestions for new physics
at the weak scale have been made, however we focus here on
supersymmetry (SUSY).

Supersymmetry is being actively searched for at the LHC and, in some
circumstances,  it may even be possible to find evidence for supergravity
at the LHC \cite{Buchmuller:2004rq,Cheung:2010mc}.  Here we
contemplate whether supergravity may be responsible for a phenomenon
which has already been observed---the existence of non-zero neutrino
masses, much smaller than the masses of the charged leptons. 

Models of neutrino masses abound, and can be broadly categorized
by whether neutrino masses are Majorana or Dirac.  The majority of
such models assume lepton-number violation at some scale, which
usually implies Majorana neutrino masses.  Often these models employ
 some manifestation of the `seesaw' mechanism \cite{Minkowski:1977sc,
 Yanagida:1979as,GellMann:1980vs,Glashow:1979nm,
 PhysRevLett.44.912,Schechter:1980gr,Schechter:1981cv}.\footnote{Extending the seesaw to a SUSY framework one finds 
 examples where Majorana neutrino masses are purely supersymmetric, arise as a result of R-parity violation \cite{Aulakh:1982yn,Hall:1983id,Lee:1984kr,Lee:1984tn,Ellis:1984gi,Dawson:1985vr}, or as a 
 result of SUSY breaking at the TeV scale \cite{ArkaniHamed:2000bq,
 ArkaniHamed:2000kj,Borzumati:2000mc,Borzumati:2000ya,MarchRussell:2004uf,
 MarchRussell:2009aq}.}

Although much less prevalent, models which generate small Dirac
neutrino masses have also been proposed, and can arise in a
number of contexts, including composite neutrino models
\cite{ArkaniHamed:1998pf,Grossman:2008xb,Okui:2004xn},
supersymmetric, and supersymmetry-breaking models
\cite{Langacker:1998ut,PhysRevLett.100.091804,Marshall:2009bk,Abel:2004tt},
extra-dimensional scenarios
\cite{Grossman:1999ra,PhysRevD.65.024032,Dienes:1998sb,Hung:2002qp,Gherghetta:2003he,Davoudiasl:2005ks},
models with extra discrete or continuous gauge symmetries
\cite{Krauss:1988zc,Gogoladze:2001kj,PhysRevD.75.055009} or unparticle scenarios \cite{vonGersdorff:2008is}.
Often, new mass scales or small parameters must be introduced
in an ad hoc manner, reducing both the explanatory power of the
model and the ability to make concrete predictions.  An example of
this arises in some seesaw neutrino mass models, where the need
to introduce a large Majorana mass for right-handed neutrinos
simply trades the problem of explaining a low mass scale with that
of explaining a high mass scale.  Unless one addresses the
question of why right-handed neutrino masses take the required
values, the observed scale of neutrino masses remains mysterious.

In this work we present a new class of supersymmetric models which, under the assumption of low-scale supersymmetry breaking, TeV-mass superpartners, and $\mathcal{O}(1)$ couplings, accommodate Dirac neutrino masses of the correct scale, without the need
for unusually small couplings, or the introduction of new mass scales
beyond those already required for SUSY theories.  The first prediction
of these models is that neutrino masses are Dirac and thus
neutrino-less double $\beta$-decay will not be observed in future
experiments.  The second, much more model-specific, prediction is
that $R$-symmetric supersymmetry exists at the weak
scale.\footnote{Examples of $R$-symmetric models include
\cite{Hall:1990hq,Randall:1992cq,Lazarides:1998iq,Kurosawa:2001iq,Fox:2002bu,Nelson:2002ca,Kribs:2007ac,Rehermann:2011ax,Bertolini:2011tw}.  Scenarios with weak scale R-symmetry are also attractive from a naturalness perspective as Dirac gauginos can be heavier than in the Majorana case while preserving naturalness, and collider bounds are also weaker than in the Majorana case, allowing for lighter squarks and improving naturalness \cite{Heikinheimo:2011fk,Kribs:2012gx}.}
This $R$-symmetry protects neutrinos from obtaining weak scale
masses.  However, after SUSY-breaking the $R$-symmetry must
be broken in order to cancel the cosmological constant.  Rather
surprisingly, this leads to the generation of Dirac neutrino masses
at the desired scale, independent of the SUSY-breaking scale,
leading to a novel connection between the smallness of the
cosmological constant and small neutrino masses.

We find these models compelling for two reasons:  First, they make
concrete predictions for physics at the TeV scale and also at
$\beta$-decay experiments.  Second, the scale of neutrino masses
emerges naturally, independent of the scale of SUSY breaking, and
without the introduction of any new dimensionful parameters beyond those required in low-scale SUSY breaking scenarios.  This
last observation is actually quite general in low-scale scenarios.  If we define 
the scale at which SUSY breaking is mediated to the Supersymmetric Standard Model (SSM) as $M$, then in any
model with low-scale SUSY breaking the product of the gravitino mass
with a hard SUSY-breaking coefficient, i.e.\ $m_{3/2} F_X/M^2$, does
not depend on the SUSY breaking scale, as we can always borrow a
factor of $F_X$ from the gravitino mass and rewrite this formula as
$(F_X/M)^2 / M_P \propto (\text{TeV})^2/M_P$.  When the additional loop
factors required to relate $F_X/M$ to the TeV scale are included, this
comes out close to the scale of neutrino masses, independent of the
scale of SUSY breaking.  The scenario we present in this paper can
be thought of as a specific way of exploiting this coincidence to obtain
realistic neutrino masses.

In \Sec{sec:structure} we describe the general structure of these
models, and then in \Sec{sec:model} go on to present an explicit
model which realizes the desired features.  \Sec{sec:colliders} discusses
novel collider signatures, and \Sec{sec:conclusions}
contains brief conclusions.  Throughout, we will denote
superfields in bold, component fields by the same symbol in
plain font, and scalars with $R$-charge carry a tilde.

\section{General Structure}\label{sec:structure}

First, let us establish our assumptions.  We consider supersymmetric
models with an unbroken anomaly-free $R$-symmetry\footnote{It is
not necessary to consider a continuous $U(1)_R$ symmetry; the
discrete subgroup $\IZ_p$, for large enough $p$, has the same
implications.} at the TeV scale, which is then broken at some much
lower scale ($<~\!\!10 \GeV$) by SUGRA effects.  We assume that
supersymmetry breaking is communicated to the visible sector
by some non-gravitational mechanism, such as gauge mediation, in
order to generate sufficiently large soft masses for the standard model
superpartners.  We also assume that the left-handed neutrinos are
forbidden from obtaining Majorana masses after electroweak symmetry
breaking i.e.\ that the Weinberg operator,
$\int d^2\theta~\mb{H_u H_u L L}\,$, is absent from the
Lagrangian.\footnote{If this operator arises due to Planck scale physics
it leads to subdominant contributions to neutrino masses.  We assume
that it does not arise due to physics at lower scales as a result of gauge,
or global, symmetries.}  Finally,  we restrict the standard model fermions
and Higgs boson to have $R$-charge 0 (this is the case in most known
$R$-symmetric models, as will be discussed in \Sec{sec:model}, but not
necessarily in more exotic scenarios \cite{Frugiuele:2011mh}).

In order to avoid light charginos and cancel gauge anomalies, we include
another doublet $\mb{R_u}~\sim~(\mb{1},\mb{2},-\frac 12)$ of $R$-charge
2, which allows a weak-scale $\mu$ term,
\begin{equation} \label{eq:mu_term}
    \cL_\mu = \int d^2\th\, \mu \mb{H_u R_u} ~.
\end{equation}
If we introduce right-handed neutrino superfields $\mb{N}$, with
$R$-charge $Q_R (\mb{N}) \neq 1$ then Dirac neutrino masses are
forbidden by the $R$-symmetry, as desired.  If we consider the
particular case where $Q_R (\mb{N}) = 3$, then the most general
renormalisable coupling of the right-handed neutrinos is
\be
    \mathcal{L} \supset \l_N \tilde{R}_u^\dagger L N~ + \mathrm{h.c.} ~~.
    \label{eq:lag}
\ee
This term is not holomorphic, and must arise as a result of
SUSY-breaking in non-renormalizable \kahler potential terms.  So we see
that the lowest-dimensional term which couples the right-handed neutrinos
to the Higgs sector arises, after SUSY-breaking, from
\be
K \supset \l\frac{\mb{X^\dagger R_u^\dagger}}{M^2} \mb{L N} ~~,
\label{eq:kahler}
\ee
where $\mb{X}$ is a SUSY-breaking spurion chiral superfield with
$R$-charge 2, $M$ is the mass of some messenger fields which we expect 
to be at the scale of the gauge-mediation messengers, and $\l$ is an
unknown coefficient which is generated by integrating these messengers out.
We would expect that $\l$ is of order a loop factor.  This means that $\l_N$ is
typically small, being given by $\l_N = \l F_X/M^2$.

Now, we are assuming that neutrinos cannot gain Majorana masses, and Dirac
masses are forbidden by the $R$-symmetry.  However an additional factor, which
provides the motivation for this set-up, is that we know in SUGRA the $R$-symmetry
must be broken in order to tune away SUSY-breaking vacuum energy and allow
for a small cosmological constant.  Following many authors, we do not attempt to
explain how this occurs, but simply allow for a constant in the superpotential of
$W_0 = F_X M_P/ \sqrt{3}$.  This $R$-symmetry breaking generically leads
to soft SUSY-breaking contributions to the scalar potential of order
$m_{3/2} = F_X/\sqrt{3} M_P$.  In the context of gauge-mediated models these
extra terms are small, and therefore usually innocuous, but in this set-up they
are the only source of $R$-symmetry breaking.

Once the $R$-symmetry is broken in this way, supersymmetric terms such as
\eqref{eq:mu_term} give rise to corresponding SUSY-breaking $B$-terms in
the scalar potential, which violate the $R$-symmetry.  These can be
calculated in a number of ways, however the conformal compensator
formalism \cite{Cremmer:1978hn, Cremmer:1982en,Luty:2000ec} is
most direct.  Taking the superconformal compensator superfield to be
$\mb{\phi}$, with $\langle \mb{\phi} \rangle = 1 + \theta^2 m_{3/2}$, then
the superpotential $\mu$-term
\be
W \supset \mb{\phi}^3 \mu\, \mb{H_u R_u} ~~,
\ee
in combination with the usual \kahler potential terms involving $\mb{\phi}$,
leads to an $R$-symmetry-breaking $B_\mu$-term,
\begin{equation}
        \cL ~\supset~ - \mu m_{3/2} \tilde R_u H_u ~+ \mathrm{H.c.}
\end{equation}
Both $H_u$ and $\tilde{R}_u$ have gauge-mediated soft masses at the weak
scale, so this term does not destabilize the scalar potential.  However, once
electroweak symmetry is broken by the VEV of $H_u$, it gives rise to a tadpole
for $\tilde{R}_u$, and therefore an $R$-symmetry-breaking VEV.  Since the
linear term has a Planck-suppressed coefficient, we can disregard any terms of
order $(\tilde R_u)^3$ and higher in the potential, and estimate this VEV to be 
\be
\langle \tilde{R}_u \rangle \approx \frac{1}{\sqrt{2}} m_{3/2} \frac{\mu v_u}{m_{\tilde{R}_u}^2} ~~,
\ee where $m_{\tilde{R}_u}$ is the total mass of $\tilde{R}_u$, including supersymmetric and non-supersymmetric contributions.


The up-shot is that, due to the $R$-symmetry breaking required to cancel the
cosmological constant, $\tilde{R}_u$ obtains a small VEV, and therefore generates
non-zero neutrino masses through the interaction \eqref{eq:kahler}.  If we assume
that $\mu \approx v_u \approx m_{\tilde R_u}$, their scale is
\be
m_\nu \approx \l \frac{F_X}{\sqrt{2}M^2} m_{3/2}  ~~.
\ee

Na\"ively then, it seems that we obtain neutrino masses proportional to the
gravitino mass.  However, if we use $m_{3/2} = F_X/\sqrt{3} M_P$ to replace
$m_{3/2}$ in the above, we get
\begin{equation}\label{eq:neutrino_mass}
    m_\nu \approx \l \frac{F_X^2}{\sqrt{6}M^2} \frac{1}{M_P} ~.
\end{equation}
Since we are working within the framework of gauge-mediation, soft masses
come dressed with a loop factor, and are given schematically by
\begin{equation*}
    \tilde{m} \sim \frac{\alpha}{4\pi} \left | \frac{F_X}{M_M} \right | ~,
\end{equation*}
where $M_M$ is the mass of the gauge-mediation messengers, satisfying $M_M\sim M$,\footnote{We
are allowing for $M_M\neq M$, since the gauge mediation messengers need
not have exactly the same mass as the messengers generating
\eqref{eq:kahler}.}  and $\a$ is the relevant fine structure constant.  For
squarks, the strong interaction dominates, and we can take $\a = \a_s$.  If
we demand that squark masses are at $\sim 1 \TeV$, then
\Eq{eq:neutrino_mass} becomes
\begin{equation} \label{eq:miracle}
    \begin{split}
        m_\nu & \sim~ \frac{\l (16 \pi^2)}{\sqrt{6}\a_s^2} \left(\frac{M_M}{M}\right)^2 \frac{(\text{TeV})^2}{M_P}\\ 
        & \sim~ 2.2\, \l \left(\frac{M_M}{M}\right)^2 \text{ eV}  ~~,
    \end{split}
\end{equation}
where we have used the value of $\a_s$ at the $Z$ pole, $\a_s \approx 0.11$.
We see here the coincidence of scales mentioned in \Sec{sec:introduction};
this expression compares favorably to the experimental data, which
tells us that the largest neutrino mass satisfies \cite{Nakamura:2010zzi}
\begin{equation*}
    0.04~\mathrm{eV} \lesssim m_\nu \lesssim 1.7~\mathrm{eV} ~.
\end{equation*}
We emphasize again that we have not introduced any new mass scale in
the theory beyond those relevant in any gauge-mediated SUSY model, namely the TeV scale
and the Planck scale.  The overall scale of SUSY-breaking has not been set,
and remains a free parameter, the only assumption in this regard is that it is small
enough for gravity-mediated effects to be sub-dominant.  This is not a serious
constraint, and in fact $m_{3/2}$ could be as large as $10$ GeV or so.  We also assume that $R$-symmetry breaking originates
entirely as a result of canceling the cosmological constant.

In the next section we put these considerations on a firmer footing by
building an explicit model that generates the interaction in \Eq{eq:kahler}.

\section{Concrete Models}\label{sec:model}

Our discussion so far has been framed in quite general terms, so we will
now demonstrate how the scenario we have outlined can arise in a
specific model.  In fact, it can be equally well implemented in the
`Minimal $R$-Symmetric Supersymmetric Standard Model' (MRSSM)
\cite{Kribs:2007ac}, and the `Supersymmetric One Higgs Doublet Model'
(SOHDM) \cite{Davies:2011mp}.  In these models the $R$-charge
assignments are as detailed in \Tab{tab:spectrum}, and we note that the
$R$-symmetry is non-anomalous with respect to the standard model gauge
group.   To either model, we add an additional right-handed neutrino
superfield $\mb{N}$, with $Q_R (\mb{N}) = 3$.  Note that the
notation $\mb{R_u}$ for the $R$-charge 2 doublet of \Sec{sec:structure}
is borrowed from the MRSSM, while in the SOHDM, $\mb{R_u}$ can be
identified with the field which was called $\mb{\eta}$ in \cite{Davies:2011mp}.
\begin{table}[htp]
\begin{center}
\begin{tabular}{| l | l | c | r | r |}
\hline
    ~Field ~&~ Gauge rep. ~&~ $R$-charge ~~ \\
\hline
    ~~$\mb{L}$ & ~~$(\rep{1},\rep{2}, -\frac 12)$ & $1$\\

    ~~$\mb{E^c_i}$ & ~~$(\rep{1},\rep{1}, 1)$ & $1$\\

    ~~$\mb{H_u}$ & ~~$(\rep{1},\rep{2}, \frac 12)$ & $0$\\

    ~~$\textcolor{red}{\bm{H_d}}$ & ~~$\textcolor{red}{(\rep{1},\rep{2}, -\frac 12)}$ & $\textcolor{red}{0}$\\

    ~~$\mb{T}$ & ~~$(\rep{1},\rep{3}, 0)$ & $0$ \\
    
    ~~$\mb{S}$ & ~~$(\rep{1},\rep{1},0)$ & $0$ \\

    ~~$\mb{X}$ & ~~$(\rep{1},\rep{1},0)$ & $2$ \\

    ~~$\mb{W'_\alpha}$ & ~~$(\rep{1},\rep{1},0)$ & $1$ \\
    
     ~~$\bm{R_u}$ & ~~$(\rep{1},\rep{2}, -\frac 12)$ & $2$\\

    ~~$\textcolor{red}{\bm{R_d}}$ & ~~$\textcolor{red}{(\rep{1},\rep{2}, \frac 12)}$ & $\textcolor{red}{2}$\\
    
    ~~$\mb{N}$ & ~~$(\rep{1},\rep{1}, 0)$ & $3$ \\
\hline
\end{tabular}
{\caption{\label{tab:spectrum}
The relevant chiral superfield content of the $R$-symmetric models.  The
content of the SOHDM is denoted in black (with a notational change,
$\mb{\eta} \to \mb{R_u}$, relative to \cite{Davies:2011mp}), and additional
fields and $R$-charges required for the MRSSM are denoted in red.  Gauge
superfields and colored superfields are not shown.  The fields
$\mb{X}$ and $\mb{W'_\alpha}$ are the spurion superfields parametrising SUSY
breaking.}}
\end{center}
\end{table}

In order to construct a complete model we must specify a messenger
sector which mediates SUSY-breaking terms with the structure we
desire.  There exist a number of studies of $R$-symmetric gauge-mediation
(see e.g.\ \cite{Fox:2002bu,Antoniadis:2006uj,Amigo:2008rc,Blechman:2009if,Benakli:2008pg,Benakli:2009mk,Benakli:2010gi,Carpenter:2010as,Abel:2011dc}),
and we will assume the presence of, for example, the gauge-mediation
messengers of  \cite{Benakli:2010gi}, with mass $M_M$.  In order to generate
the term \eqref{eq:kahler}, we need to introduce further vector-like matter,
\begin{align*}
    \bm{A}'_i \sim (\rep{1},\rep{1},0)_{-1} & ~~,~~ \bm{\overline{A}}'_i  \sim (\rep{1},\rep{1},0)_0 \\
     \bm{B}'_i \sim  (\rep{1}, \rep{1}, 0)_{4} & ~~,~~  \bm{\overline{B}}'_i  \sim (\rep{1}, \rep{1}, 0)_{-2}\\
     \bm{L}' \sim  (\rep{1},\rep{2},-\frac 12)_3 & ~~,~~  \bm{\overline L}'  \sim (\rep{1},\rep{2},\frac 12)_{-1} \\
         \bm{N}'  \sim  (&\rep{1},\rep{1},0)_1 
\end{align*}
where $i$ is the generation index, and the subscripts are the
$R$-charges. Notice that the $R$-symmetry remains non-anomalous, and
the pair $\bm{L}'$ and $\bm{\overline L}'$ could also play the role of
gauge-mediation messengers.\footnote{Unification is spoiled with the addition of this matter content to the MSSM, however, MSSM unification predictions have already been lost through the addition of the Dirac gaugino partners.}
The symmetries of this model allow the superpotential
\begin{equation}
    \begin{split}
        W  ~=  &~~  M (\bm{A}'_i\bm{\overline{A}}'_i + \bm{B}'_i\bm{\overline{B}}'_i + \bm{L}'\bm{\overline L}' 
            + \frac{1}{2} \bm{N}'\bm{N}' )\\
        & + \l_{ij}' \bm{X A}'_i \bm{\overline{B}}'_j + \l_A \bm{L}_i \mb{A}'_i \bm{\overline L}' \\
         &   + \l_B \bm{N}_i  \bm{N}' \mb{\overline{B}}'_i + \l_R \bm{R_u} \bm{N}' \bm{\overline L}'  ~~,
    \end{split}
\end{equation}
where we have assumed the same mass for all of the extra matter fields
for simplicity.  We are also assuming a standard $\IZ_2$ `messenger parity'
in order that messenger sector fields do not mix with SM fields.\footnote{If
this symmetry were absent then $\mb{N}'$ could lead to Majorana neutrino
masses.}  This is an $R$-symmetric analogue of a model from
\cite{Ibe:2010ig} used to generate non-supersymmetric down-type quark
Yukawa couplings, and indeed, integrating out the extra matter in this model
generates the \kahler term of \Eq{eq:kahler} at one loop.\footnote{These
couplings also generate TeV-scale soft masses for the right-handed
sneutrinos.  In an expansion in $F_X/M$ this mass is
${\tilde{m}_{N}^2} = \frac{|\l_B|^2 {\l'}^\dagger \l' }{48 \pi^2} \left| \frac{F_X}{M} \right|^2$, and the right-handed sneutrino flavor structure is aligned with that of the neutrinos.  Additional flavor off-diagonal terms are also generated for the left-handed sleptons, $\D{\tilde{m}_{L}^2} = \frac{|\l_A|^2 \l' {\l'}^\dagger}{48 \pi^2} \left| \frac{F_X}{M} \right|^2$.  After diagonalizing lepton masses this introduces factors of the PMNS matrix in lepton-slepton-chargino vertices.  This satisfies lepton MFV, and even without the MFV structure, large off-diagonal left-handed slepton mixings are allowed in $R$-symmetric scenarios \cite{Fok:2010vk}.}
Including the generation structure and couplings, to first order in $\mb{X}^\dagger/M^2$, this term becomes
\be
    K \supset \frac{{\l'}^\dagger_{ij} \l_R^\dagger \l_A \l_B }{48 \pi^2} \frac{\mb{X^\dagger R_u^\dagger}}{M^2} \mb{L_i N_j} ~~.
    \label{eq:kahler2}
\ee
Comparing to \Eq{eq:kahler}, we can read off the value of $\l$ in this model;
substituting this into \Eq{eq:miracle}, we find neutrino mass parameters of
\begin{equation}
    \begin{split}
        m_{\nu,ij} & \sim~ \frac{\l'_{ij} \l_A \l_B \l_R}{3\sqrt{6}\a_s^2} \left(\frac{M_M}{M}\right)^2 \frac{(\text{TeV})^2}{M_P} \\
        & \sim~ 0.005\, \l'_{ij} \l_A \l_B \l_R \left( \frac{M_M}{M} \right)^2 \left( \frac{\tilde{m}_{\tilde{Q}}}{\text{TeV}} \right)^2  \text{ eV} ~~.
    \end{split}
\end{equation}
Thus for couplings $\l \sim \mathcal{O}(1)$, and messenger masses of
$M_M \sim 3 M$,\footnote{Or, alternatively heavier squarks or larger
couplings.} neutrino masses come out just right for TeV-scale squarks.

It should be noted that in the case of the MRSSM model we could have instead taken $\mb{N}_i$ to have $Q_R =-1$.  In that case holomorphic superpotential terms such as $W \supset \mb{R}_d \mb{L}_i \mb{N}_i$ would have been allowed.  In this scenario neutrino masses would come out at $m_{\nu} \sim m_{3/2}$ \cite{Rehermann:2011ax}.  However this would require very low-scale SUSY breaking, and neutrino masses would not be independent of the SUSY-breaking scale.

\section{Collider Signatures}\label{sec:colliders}
\begin{figure}[htp]
\centering
\includegraphics[height=2.2in]{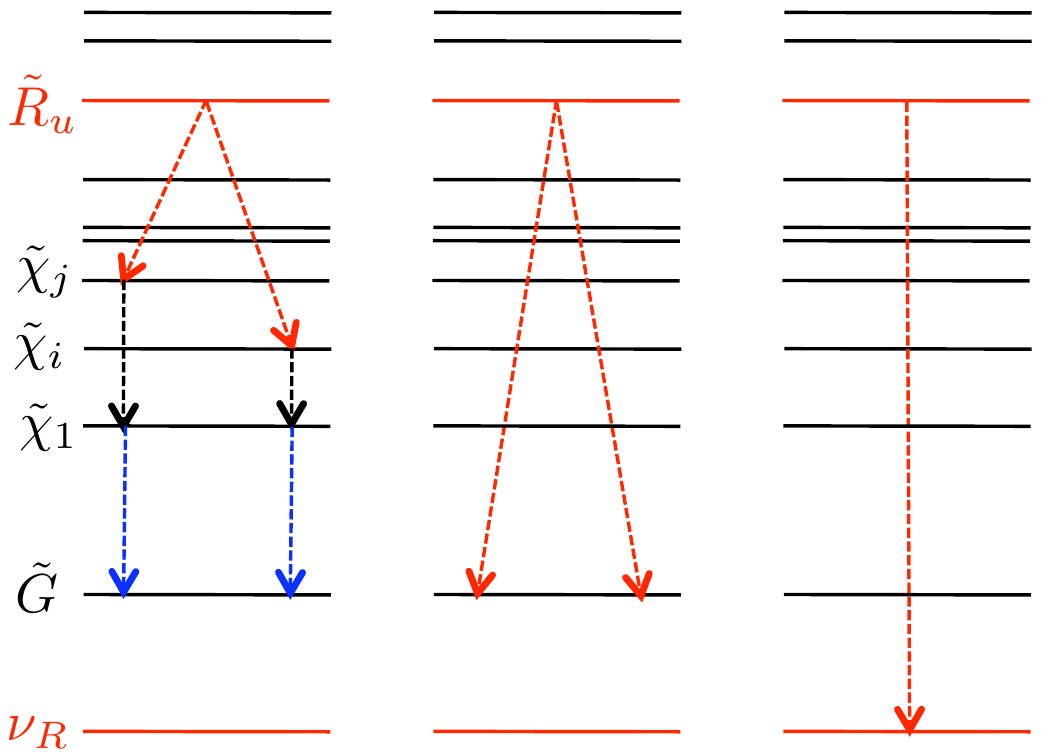}
\caption{Possible decay chains for a single $\tilde{R}_u$.  $Q_R =1$ SSM sparticles are denoted in black.  SSM superpartners carry the label $\tilde{\chi}_i$.  $Q_R =2$ particles are denoted in red.  On the left, if kinematically allowed, each $\tilde{R}_u$ decays to a pair of $Q_R =1$ sparticles, which each then decay to the LOSP, which finally decays to the gravitino.  The rates for these processes depend on the particle spectrum and SUSY-breaking scale.  This first possibility is interesting as each decay likely involves the emission of SM particles and hence the process $\tilde{R}_u^\dagger \tilde{R}_u \rightarrow 4 \tilde{\chi}_i \rightarrow 4 \tilde{\chi}_0\rightarrow 4 \tilde{G}$ results in very high multiplicity final states.  In the middle we depict the direct decay to gravitinos, $\tilde{R}_u^\dagger \tilde{R}_u \rightarrow 4 \tilde{G}$.  On the right we show the direct decay to leptons, $\tilde{R}_u^\dagger \tilde{R}_u \rightarrow U_{ij} U_{kl}^\dagger  \nu_{R_i} l_{L_j}  \overline{\nu}_{R_k} \overline{l}_{L_l}$.  This process is particularly striking as it results in the pair production of isolated leptons with flavor determined by the PMNS matrix, $U$.}
\label{fig:decays}
\end{figure}

The models discussed possess a number of interesting collider signatures.  For example, the connection between neutrino mass generation and SUSY breaking leads to non-degenerate slepton masses.  Arguably the most interesting signatures arise due to the $R$-symmetric structure.  The SUSY phenomenology of the $Q_R = 1$ particles largely resembles that of the R-parity odd particles in a gauge-mediated version of the SSM, albeit with non-trivial squark and slepton flavor structure allowed.  However the $Q_R =2$ scalar doublet $\tilde{R}_u$ possesses novel decay signatures due to the $R$-symmetry.  The charged and neutral scalars can be produced in proton-proton collisions through Drell-Yan processes, and for $m_{\tilde{R}_u} \approx 200$ GeV, the pair production cross-section at the $7$ TeV LHC is of order $\sigma \sim 15$ fb.  $R$-symmetry violating decays of $\tilde{R}_u$ are allowed due to the SUGRA effects, however these are controlled by the ratio of $m_{3/2}$ to the TeV-scale and so we expect these decays to be subdominant to $R$-symmetry preserving channels.  Hence the decay of each $\tilde{R}_u$ must result in one $Q_R =2$ particle or two $Q_R =1$ particles.  Some of the possible decay modes are depicted in \Fig{fig:decays}.

Although a full study of the collider phenomenology of these models is beyond the scope of this work, we note some interesting features here.  The decays to high multiplicity final states are interesting in their own right, however the direct leptonic channel has the intriguing feature that leptonic branching ratios are determined by the PMNS matrix, $U$, and so we focus on this case.

\begin{figure}[htp]
\centering
\includegraphics[height=1.27in]{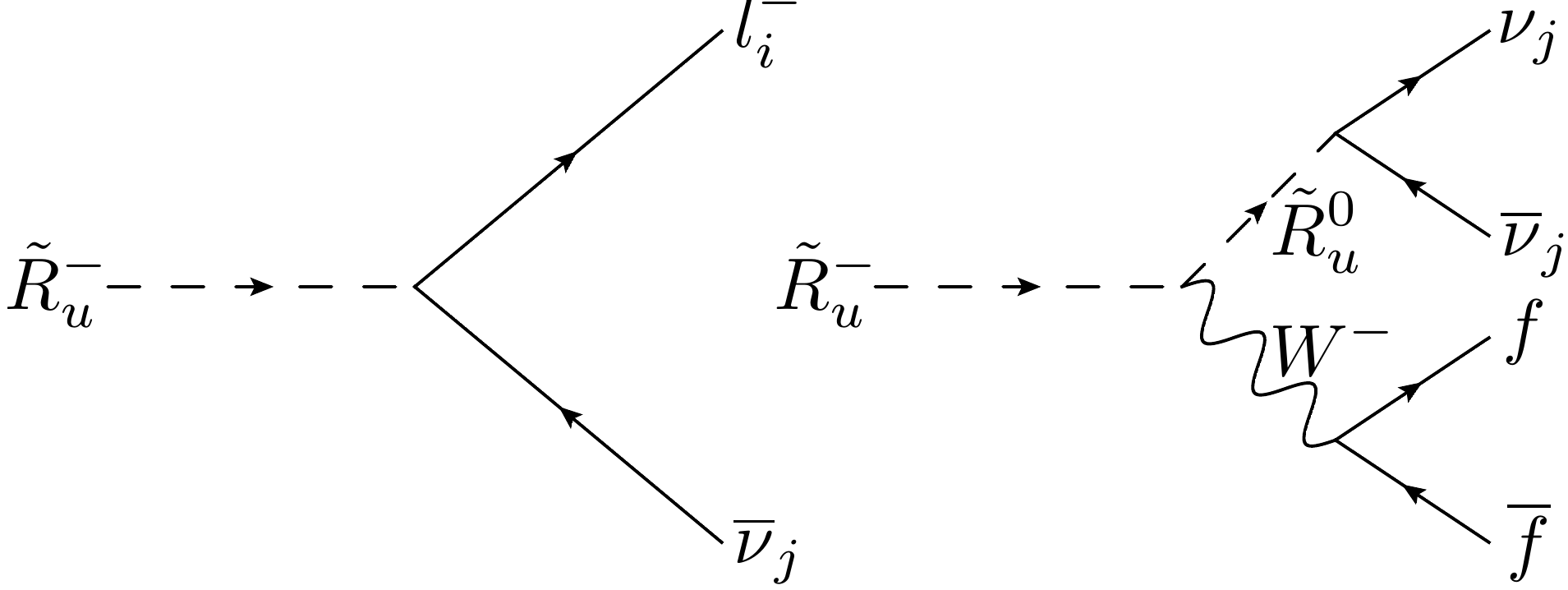}
\caption{Decay modes of the $R$-charge 2 scalar which terminate with an $R$-charge 2 neutrino.  The direct decay to a neutrino and charged lepton dominates for $m_{\nu_{3}}/m_{3/2} \gtrsim 10^{-3}$ in the case of the normal hierarchy.  In this case the pair production $\tilde{R}_u^\pm$ would have collider signatures very similar to $W^\pm$ pair production, however the final state would be purely leptonic and would not be flavor-blind, as it is determined by the flavor composition of the heaviest neutrino.  In the case of the normal hierarchy this gives roughly equal number of $\mu$ and $\tau$ final states and negligible $e$ for small $\theta_{13}$.}
\label{fig:decaysfeyn}
\end{figure}

The Yukawa coupling in \Eq{eq:lag} allows for the direct decay $\tilde{R}_u^0 \rightarrow \nu_L \nu_R$, however this decay will be invisible.  The charged state can decay  $\tilde{R}_u^- \rightarrow \nu_R l^-_L$ however SUSY dictates that this state has mass $m_{\tilde{R}_u^-}^2 = m_{\tilde{R}_u^0}^2 + M_W^2$ and so can also decay via a virtual $W$ emission to $\tilde{R}_u^0$ which will then decay invisibly.  Both decay modes are depicted in \Fig{fig:decaysfeyn}.  We find that as long as the neutrino Yukawa couplings aren't too small, (i.e\ $m_{\nu_{3}}/m_{3/2} \gtrsim 10^{-3}$ as would automatically be satisfied in the SOHDM \cite{Davies:2011mp} setup), then the direct decay dominates.  This decay mode has the intriguing property that it is determined by the neutrino mass matrix, and hence $\tilde{R}_u^-$ couples dominantly to the heaviest neutrino species.  In the case of the normal hierarchy this picks out decays to $\nu_3$ and hence the flavor of the charged lepton observed in these decays is determined by the third column of the PMNS matrix.  This implies that decays will be dominantly to $\mu$ and $\tau$ but not $e$ in the case of the normal hierarchy.  For the inverted hierarchy the decays will be mostly to the two heaviest neutrinos and one would expect an excess in $e$ compared to $\mu$ or $\tau$.

That the neutrino mixing angles could in principle be `measured' at the LHC through $\tilde{R}_u$ decays is exciting, however the full parameter space is large, and the correct treatment of backgrounds is complicated, so we leave a full collider study of the multi-lepton signatures of this class of models to future work.

\section{Conclusions}\label{sec:conclusions}
Supersymmetry is a well-motivated solution to the hierarchy problem and the LHC has begun to explore the scales at which we expect it to become apparent.  As yet, no signs of SUSY have been observed.  While the first signals of SUSY might well be observed in leptonic channels at the LHC, it is interesting to consider whether the basic ingredients of SUSY theories could address a major puzzle in the neutrino sector, namely the existence of non-zero neutrino masses.  In this work we have described a class of supersymmetric models wherein non-zero neutrino masses arise at the desired scale, independent of the overall scale of SUSY breaking.  This relies on low-scale mediation of SUSY breaking, such as gauge-mediation, and occurs as a result of the $R$-symmetry breaking necessary to tune the cosmological constant to small values.

These models make two testable predictions: neutrino-less double beta-decay should not occur in nature and $R$-symmetric SUSY should exist at the TeV scale.  In addition, non-degenerate left-handed sleptons arise as a general feature of these models and charged scalar decays at colliders can lead to isolated leptons, with flavor structure determined by the PMNS matrix.

\section{Acknowledgements}
MM gratefully thanks Jesse Thaler and Keith Rehermann for discussions.  R.D. is supported by the Engineering and Physical Sciences Research Council [grant number EP/H02672X/1] and MM is supported by a Simons Postdoctoral Fellowship.  

\newpage
\bibliography{Rneutrinosref}

\end{document}